\documentclass[prb,twocolumn,superscriptaddress,preprintnumbers,amsmath,amssymb,floatfix]{revtex4}
\usepackage{graphicx}

\begin{document}
\title{Electronic structure and magnetic properties 
of metallocene  multiple-decker sandwich nanowires}
\author{C. Morari}
\affiliation{National Institute for Research and Development of Isotopic and 
Molecular Technologies, 65-103 Donath,  RO-400293 Cluj Napoca, Romania}
\author{H. Allmaier}
\affiliation{Institute of Theoretical Physics, Graz University of Technology,
A-8010 Graz, Austria}
\author{F. Beiu\c seanu}
\affiliation{Department of Physics, University of Oradea, RO-410087 Oradea, Romania}
\author{T. Jurcu\c t}
\affiliation{Department of Physics, University of Oradea, RO-410087 Oradea, Romania}
\author{L. Chioncel}
\affiliation{Augsburg Center for Innovative Technologies, University of Augsburg, D-86135 Augsburg, Germany}
\affiliation{Theoretical Physics III, Center for Electronic Correlations and Magnetism, Institute of Physics,
University of Augsburg, D-86135 Augsburg, Germany}

\begin{abstract}
We present a study of the electronic and magnetic properties of the multiple-decker sandwich 
nanowires ($CP-M$) composed of cyclopentadienyl ($CP$) rings and $3d$ transition metal atoms 
(M=Ti to Ni) using first-principles techniques. We demonstrate using Density Functional Theory 
that structural relaxation play an important role in determining the magnetic ground-state 
of the system. Notably, the computed magnetic moment is zero in $CP-Mn$, while in $CP-V$ a 
significant turn-up in magnetic moment is evidenced. Two compounds show a half-metallic 
ferromagnetic ground state $CP-Fe/Cr$ with a gap within minority/majority spin channel. In 
order to study the effect of electronic correlations upon the half-metallic ground states 
in $CP-Cr$, we introduce a simplified three-bands Hubbard model which is solved within the 
Variational Cluster Approach. We discuss the results as a function of size of the reference cluster 
and the strength of average Coulomb $U$ and exchange $J$ parameters. Our results demonstrate 
that for the range of studied parameters $U=2-4eV$ and $J=0.6-1.2eV$ the half-metallic character 
is not maintained in the presence of local Coulomb interactions.
\end{abstract}


\maketitle

\section{Introduction}
Utilizing the spin degree of freedom of electrons in the solid state
electronics has led to the emergence of the rapidly developing field 
of spintronics \cite{zu.fa.04}. Important electronic applications based on the magnetoresitive 
effect in two dimensional heterostructures already exist. Typical devices are 
magnetic read heads and non-volatile magnetic random access memories, where the relative 
alignment of the layer magnetizations causes large variations of the resistance
of the structure. This effect is known as giant magnetoresistance \cite{bi.gr.89}, and was 
discovered in $FeCr$ multilayer \cite{ba.br.88}, in which the magnetization of the 
layers couples by the indirect exchange interactions, mediated by the electrons of the 
intermediate layer \cite{brun.95}. 

There are continuing efforts to improve materials fabrications and design 
devices for layered magnetic structures. With the development of nanotechnology, 
quantum structures with dimension of the order of molecular or atomic 
size becomes accessible.  
In this context, the first-principle studies of atomic chain structures, 
that can produce high polarization effects are important. For instance, complete 
spin-polarization in the absence of magnetic field was predicted for several 
bulk materials in the class of half-metallic ferromagnets~\cite{gr.mu.83, ka.ir.08}.
The same properties were predicted for organic nanowires, such as $(C_6H_6)_2 V$ 
\cite{ma.ba.06} in  $(C_5H_5)Fe$\cite{zh.ya.08b} multiple decker nanowires
or in transition metal doped silicon nanostructures \cite{le.ku.10,du.ca.10}. Organic 
materials, when used as components of spintronic devices have significant advantages 
over inorganic ones. First, the spin-orbit and hyperfine interactions are weak~\cite{sa.ro.06}, 
leading to considerably long spin relaxation length and spin-lifetime~\cite{ts.al.99, ki.aw.99}.
Organic materials are cheap, low-weight, mechanically flexible, and chemically inactive, 
therefore organic spintronics received a considerable attention, in particular because the 
spin-polarized signal can be mediated and controlled in this case by organic molecules \cite{xi.wu.04}.

Metallocene molecules with the general formula $(C_5H_5)_2M$ consist of two cyclopentadienyl 
anions ($CP$) bound to a metal center. The metal atom is typically a transition metal element 
from the middle of the $3d$-row (i.e Ti,  V, Cr, Mn, Fe, Co, Ni) as well as from the column 
of iron (i.e. Ru and Os). Molecular structures such as $CP_2V$, $CP_2Cr$, $CP_2Mn$, $CP_2Fe$, 
$CP_2Co$, $CP_2Ni$ has been studied both experimentally \cite{haal.79c,ga.ha.75,haal.79b,he.he.75,al.ga.76} 
and theoretically~\cite{xu.xi.03}. Studies were also reported on the multi-decker sandwich clusters 
based on $C_5H_5$ ligands such as triple-decker complexes $CP_3Ni_2^+$ and $CP_3Fe_2^+$ \cite{sa.we.72,schi.73}
and multiple decker sandwich complexes of type $TM(CP_2Fe)_{n+1}$ with $TM$=V, Ti and $n=1-3$
\cite{na.ka.00}. The lack of individual bonds between the carbon atoms of the $CP$ ring and the metal ion
makes the $CP$ rings to rotate freely about the symmetry  axis of the molecule. Although
the MT's equilibrium structure has a D$_{5d}$ symmetry, the D$_{5h}$ has almost the same
probability to occur \cite{co.ha.06}. Within the equilibrium  D$_{5d}$ symmetry,
$d$ orbitals are splited into a $d_{z^2}$  and  two sets
of doubly degenerate orbitals $d_{xy}$/$d_{x^2-y^2}$ and $d_{xz}$/$d_{yz}$, respectively.

\begin{figure}[h]
\includegraphics[width=0.8\columnwidth]{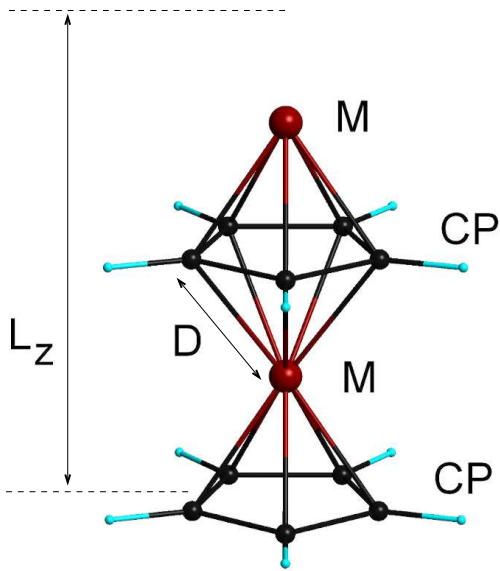}
\caption{(color online) Geometrical structure of the double decker  $(C_5H_5)_2M_2$  sandwich.
For the tetragonal unit cell we used fixed values for $L_X=L_Y=$12 \AA\, while
the values of $L_Z$ are transition metal M-dependent. $D$ illustrates the
distance between carbon atom and the transition metal atom.}
\label{struc}
\end{figure}

The electronic structure calculations based on Density Functional Theory 
predict for some metallocene wires a half-metallic ferromagnetic ground state
~\cite{zh.ya.08b,wa.ca.08}, stimulating further the search for such candidate 
materials for spintronic applications~\cite{ma.ba.06, ko.pa.07, zh.ya.08b, wa.ca.08, ku.hi.99, mi.mu.02}.
For all transition metal elements the metallocene wires share the same shape depicted in Fig.~\ref{struc}. 
It is expected that the $p-d$ covalent interactions that take place in metallocene wires
are essential for the physical and chemical properties of these compounds. Having similar 
structure, differences in the electronic properties of $(C_5H_5)_2M_2$ can be traced to 
the different electronic filling of the $d$ orbitals that determines different magnetic 
interactions. As $d$ orbitals form relatively narrow bands, in which electronic correlations 
are essential it is natural to raise the question concerning the role of electron-electron 
interactions in such compounds. 
 
In the present paper we explore the magnetic properties of 
organometallic multi-decker sandwich wires of type $(C_5H_5)M$ where $M$ is 
a transition metal with atomic numbers ranging from $Z=22$ (Ti) to $Z=28$ (Ni). 
For the interesting case of $Cp-Cr$ in which half-metalicity is predicted by Density Functional Theory-based 
calculations we formulate a low energy three-band Hubbard Hamiltonian which is solved using many-body 
techniques. Our main interest is to understand whether
half-metallic magnetism is stable against the electronic correlations in this compound. 

\section{Computational details}
We have calculated the electronic and structural properties of the $CP-M$ wires 
within the Density Functional Theory (DFT) using the SIESTA code~\cite{so.ar.02}, 
that employs pseudopotentials,
and expands the wave functions of valence electrons by linear combinations of 
numerical atomic orbitals. We have used a double-zeta polarized basis set with 
an energy cutoff of 30 meV. We have also used a number of 16 k-points to model 
the periodicity along the axis of the wire and a 2$\times$2 grid in the transverse 
direction. All calculations were performed by using generalized gradient
approximation to the exchange and correlation functional, as parameterized by 
Perdew, Burke and Ernzerhof \cite{PBE-96}.
In order to investigate the magnetism in these compounds we consider a unit
cell as shown in Fig. \ref{struc} that would naturally allow to study the 
stability of ferro vs. anti-ferromagnetic states. For all systems, we explicitly 
performed the calculations to look for the stable ferromagnetic and antiferromagnetic
configurations allowing structural relaxation. 
The atomic coordinates of all atoms in the unit cell, and the cell's length 
were relaxed up to a maximum force of  0.01 eV/\AA. The values of magnetic 
moments where computed by integrating the spin resolved charge density over 
``muffin-tin'' spheres centered on the metal atoms. The radius of the sphere was taken 
2.25 Bohr , representing about 55 - 60\% of the bond-length M-C for all complexes.
All the calculations were done using the eclipsed configuration of the
wires (i.e. $D_{5h}$ symmetry).
In order to study the stability of the wires, we compute the binding energy of the metal 
and $CP$ unit, $\Delta E$, defined as: $\Delta E = E_{wire} - 2(E_{CP} + E_M )$, where the 
total energy of the atom in the unit cell, $E_{wire}$, and that of cyclopentane 
unit, $E_{CP}$, are computed for the relaxed structures and $E_M$ is the total energy of the 
metallic atom. 

\section{Electronic structure}
Selected parameters of the relaxed wires  are given in Table \ref{tabel}. 
We comment in the followings upon the atomic-number dependence of the metal-carbon 
bond-lengths. We found that the metal - carbon bond-lengths in the investigated $CP-M$ wires 
display a regular trend. Large values are realized for the compounds with extreme atomic numbers 
(i.e. $2.29 $ \AA\ in $CP-Ti$ chain and $2.21$  \AA\ in $CP-Ni$, respectively) while the minimum
occurs at the intermediate atomic number (i.e.  $2.08 \AA$ in the  $CP-Mn$ chain).  
We note that the same trend occurs for the relaxed value of the total length, $L_Z$. 
The comparison of the computed carbon - metal bondlength shows differences ranging between 
0.5 \% (for $CP-Ni$ ) and about 3 \% (for $CP-V$). An important difference occurs for 
$CP-Mn$ (i.e. 15\%). We note that for $CP-Mn$ we also found a ferromagnetic state
which is less stable than the non-magnetic one (energy difference is 0.15 eV); for this
state the C-Mn bondlength is 2.12  \AA. This magnitude of the bond-length is in agreement with
the experimentally reported value determined for a high spin state (see Ref.~\onlinecite{haal.79c}).
The vanishing of magnetic moment in the periodic structure is probably the consequence of the large 
differences between the geometric structure of the molecule and the corresponding chain. 
\begin{table}[h]
\begin{tabular}{ccccccc}
\hline
Atom  & $L_Z$ & $D$    & $M_0$       & $M_{1/2}$  & Type & $\Delta E$   \\ 
      & [\AA] & [\AA]  &  [$\mu_B$]  & [$\mu_B$]  &  & [eV] \\ 
\hline
\hline
Ni($3d^84s^2$) & 7.35 & 2.21 (2.20$^a$)& 0.00 & 0/0        & NM  & -6.91 \\   
Co($3d^74s^2$) & 7.10 & 2.17 (2.12$^b$)& 0.00 & 1.31/-1.31 & AFM & -11.38 \\   
Fe($3d^64s^2$) & 6.77 & 2.10 (2.06$^c$)& 2.00 & 1.00/1.00  & HMF & -11.06 \\  
Mn($3d^54s^2$) & 6.59 & 2.08 (2.38$^d$)& 0.00 & 0/0        & NM  & -8.38  \\ 
Cr($3d^54s^1$) & 6.90 & 2.14 (2.17$^e$)& 2.00 & 1.00/1.00  & HMF & -7.98  \\  
V($3d^34s^2$) & 7.20  & 2.21 (2.28$^f$)& 3.99 & 1.70/1.70  & FM  & -13.73 \\  
Ti($3d^24s^2$) & 7.52 & 2.29 (2.36$^g$)& 0.00 & 1.36/-1.36 & AFM & -14.98 \\   
\hline
\end{tabular}
\caption{Geometric and magnetic properties of the wires. $L_Z$ is the length of
the unit cell, $D$ is the metal - carbon distance (see also Fig. \ref{struc}).
$M_0$ is the magnetic moment per cell, $M_{1/2}$ are the magnetic moments for
the transition metal atoms. AFM/FM - antiferro/ferro-magnetic; NM - non-magnetic,
HMF - halfmetallic ferromagnetic $\Delta E$ is the binding energy of the chain. 
Experimental data on isolated molecules are indicated for the metal-carbon bondlength.
($^a$ Reference \cite{he.he.70}, $^b$ Reference \cite{he.he.75}, $^c$ Reference  \cite{ha.ni.68}, 
$^d$ Reference, \cite{haal.79c} and
\cite{haal.75}, $^e$ and $^f$ Reference \cite{ga.ha.75}, $^g$  Reference \cite{tr.an.92})
}
\label{tabel}
\end{table}

As one can see in Tab. \ref{tabel} the values for the binding energy $\Delta E$, 
depend on the gradual filling of the d-orbitals of the metal atom as was observed 
previously~\cite{xu.xi.03}. The strongest binding energy is realized for 
$CP-Ti$; further filling the d-orbital reduces the strength of binding (i.e. non-bonding 
states are populated \cite{xu.xi.03}). An almost twice weaker binding energy is obtained 
for $CP-Cr$ (-7.98eV) in comparison with $Cp-Ti$(-14.98eV). Increasing the number  
of $d$ electrons starting from $Cr \rightarrow Mn \rightarrow Fe \rightarrow Co$ the 
binding energy is increasing again, while at $CP-Co$, the partial filling of the 
anti-bonding orbitals stops this trend. A further increase of the electronic population in the 
anti-bonding orbitals of nanowires, lead to an important reduction of the binding 
energy as seen for $CP-Ni$ wire.

\begin{figure}[h]
\includegraphics[width=0.99\columnwidth]{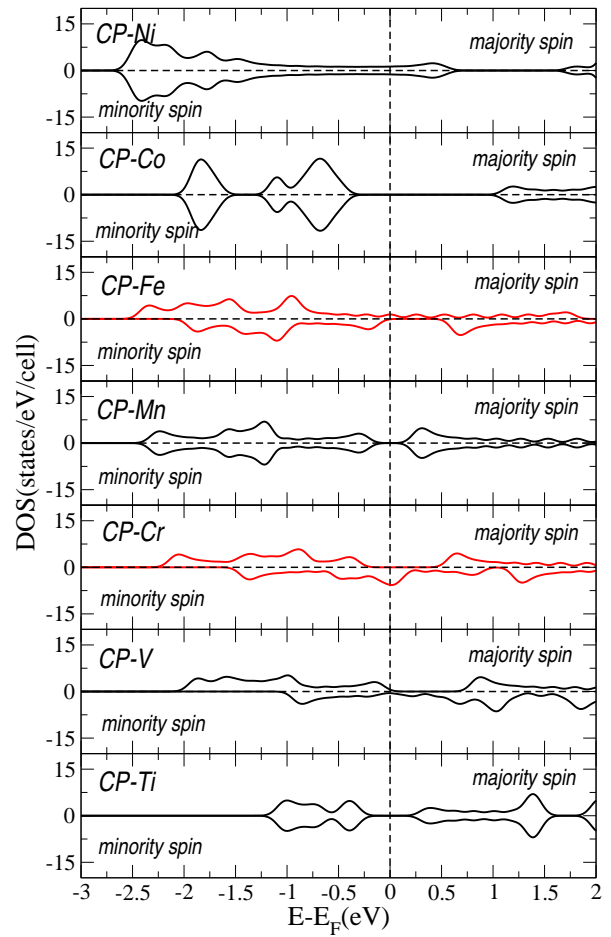}
\caption{(color online) Total DOS for the computed wires. With red line we show the cases when
half-metalicity occurs for the $CP-Fe$ (minority spin gap) and $CP-Cr$ (majority spin gap) sandwiches.}
\label{dos_omogen}
\end{figure}
Total densities of states for the seven wires are given in Fig. \ref{dos_omogen}.
Increasing the number of valence electrons from Ti ($Z=22$) to Fe ($Z=26$) results in a gradual
filling of the $d$ band, therefore the Fermi level is expected to be shifted
towards higher energies, in addition to which the magnetic interactions contribute
in determining the magnetic ground state. 
In $CP-Ti$ and $CP-V$  we found that both ferromagnetic and antiferromagnetic states
are possible. For Ti the ground-state AF configuration is lower in energy by 0.34eV, 
while for V the ferromagnetic is most stable with an energy difference of about 0.17eV.
$CP-Mn$ shows long-range ferromagnetic order for a non-relaxed 
structure that disappears in favor of a non-magnetic ground state in the relaxed case.
For all the other compounds a single stable solution has been obtained, 
for the considered range of parameters and configurations. Two systems where found to 
be half-metallic: $CP-Cr$ and $CP-Fe$. For $CP-Cr$ the gap is situated in the majority 
spin DOS and it has a value of 0.48 eV, while for the case of $CP-Fe$ the 
gap occurs in the minority spin DOS with a magnitude of about 0.41 eV. 

\begin{figure}[h]
\includegraphics[width=0.99\columnwidth]{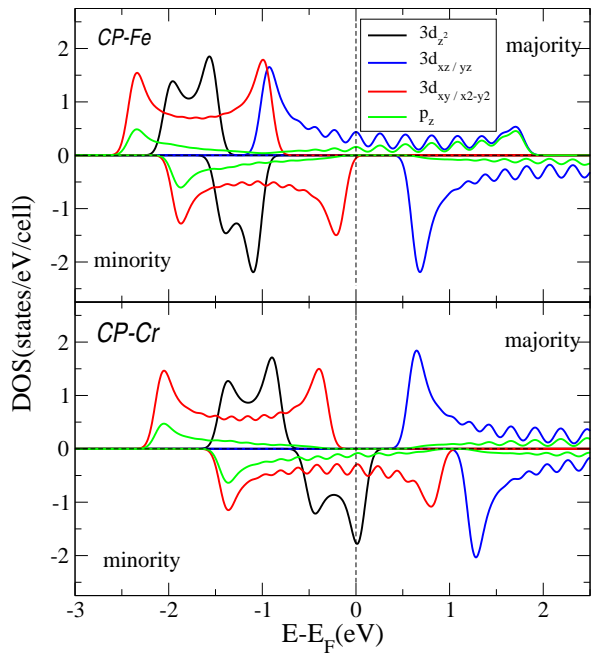}
\caption{(color online) Partial density of states for the half-metallic    
$CP-Fe$ and $CP-Cr$ wires. The degenerate M-$3d_{xz}$ and M-$3d_{yz}$
orbitals are represented with blue lines, red line represents M-$3d_{xz}$
and M-$3d_{x^2-y^2}$, and with black line the M-$3d_{z^2}$ orbitals are shown
(M=Fe,Cr).
Finally with green line we show the sum over the five C-$p_z$ orbitals.}
\label{combinate}
\end{figure}

The DOS projected onto the carbon atoms (not shown)  
has similar shapes as the total density of states but with a significantly smaller
weight. 
In order to determine the character of the orbitals near Fermi level for the 
interesting case of half-metallic $CP-Fe$ and $CP-Cr$ we computed the partial 
densities of states. In the upper panel of Fig. \ref{combinate}, the results 
for $Cp-Fe$ are seen. A significant contribution within the majority spin 
channel are provided by the degenerate $Fe-3d_{xz}/3d_{yz}$ partially occupied 
orbitals. The same orbitals are completely empty in the minority spin channel. The 
other $3d_{z^2}, 3d_{xy}/3d_{x^2-y^2}$ orbitals are completely occupied, so 
that the gap is realized between the minority spin $3d_{xy}/3d_{x^2-y^2}$ 
and $3d_{xz}/3d_{yz}$ orbitals. For the case of the majority spin half-metallic
$Cp-Cr$, the degenerate $Cr-3d_{xz}/3d_{yz}$, are empty in both majority and 
minority spin channels. Partial occupation is obtained for minority spin 
$3d_{xy}/3d_{x^2-y^2}$ and $3d_{z^2}$ orbitals, while the same orbitals are completely 
filled within the majority spin channel. Contrary to $Cp-Fe$ in $Cp-Cr$ the gap 
is realized between the majority spin $3d_{xy}/3d_{x^2-y^2}$
and $3d_{xz}/3d_{yz}$ orbitals. For both half-metals, the carbon's $p_z$ orbitals 
are also present at the Fermi level, however presenting only a weak spectral weight.
Note that, DOS projected over the $p_z$ orbitals of a carbon atom in $CP$ ring was 
 multiplied by five to facilitate the direct comparison with the total
DOS shown in Fig.\ref{dos_omogen}.
%
%
The computed magnetic moment per cell for the investigated wires 
takes values between 0 and 3.99 $\mu_B$, and can be attributed mainly to the 
transition metals, while the $CP$ rings have very weak magnetic moments.
The expected monotonic increase of the magnetic moment as function of $d$-orbitals 
occupancy can be explained by considering that the metal atom is donating one electron 
to the $CP$ ring, which forms a six electrons $\pi$ ring. Therefore, $N_{val}-1$ 
electrons rest on the metal and are determinant for its magnetic state, 
where $N_{val}$ is the number of valence electrons. By using Hund's rule and 
the transition metal valence electronic configuration (i.e.  $4s$, $3d_{z2}$, 
$3d_{xy}/3d_{x2-y2}$ and $3d_{yz}/3d_{zx}$, respectively) we can assign a 
magnetic moment to each wire. 
We expect therefore the magnetic moments of 
0,2,4,2,0 and 2 $\mu_B$ for $CP-Co$, $CP-Fe$, $CP-Mn$, $CP-Cr$, $CP-V$ and $CP-Ti$, respectively. 
Indeed, the electronic structure calculation provides these results with two remarkable exceptions: 
$CP-Mn$ (0 $\mu_B$ - instead of 4 $\mu_B$) and $CP-V$ (3.99 $\mu_B$ - instead of 0 $\mu_B$). Note
also that $Cp-Ti$ is an exception from the above discussed Hund's rule as its ground state is 
anti-ferromagnetic insulator with moments of $\pm1.36$.
According to our results, the structural relaxation effects determines the departure
from the expected behavior in all three compounds: the significant drop/turn-up of 
the magnetic moment in the $CP-Mn/V$ and the weaker reduction of the magnetic moment
in the case of $Cp-Ti$ . We have additionally checked that Mn replacing Cr in the 
frozen $CP-Cr$  structure, retains a non-zero magnetic moment (i.e. 2.36 $\mu_B$ ), 
with values close to the Cr ones (i.e. 2.0 $\mu_B$). 
An overall integer magnetic moment on the unit cell (the value $M_0$ shown in 
table Table \ref{tabel}) is obtained for all the compounds. 

\begin{figure}[h]
\includegraphics[width=0.7 \columnwidth]{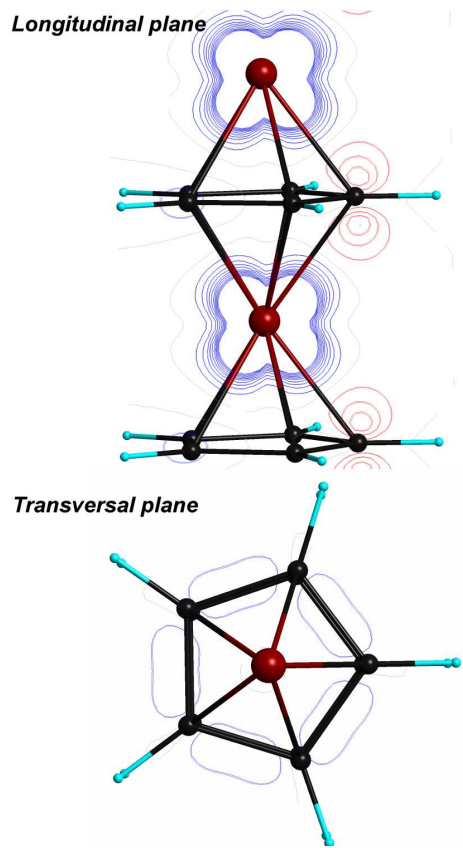}
\caption{(color online) Real-space spin density contour plots for $CP-Fe$.
Majority/minority spin density is represented with blue/red lines. The black
lines represent the regions with no difference between up and down spin densities.
Upper/lower picture presents a side/top view (see text for geometrical details of the contours).
}
\label{2Dmag}
\end{figure}

In Fig.\ref{2Dmag} we illustrate with two  contour plots the difference between majority 
and minority real space spin densities, $n({\bf r})^{\uparrow} - n({\bf r})^{\downarrow}$
for the Cp-Fe wire. We use two rectangular planes to plot the contours. 
The first plane has the edges  parallel to the $z$ axis and is passing through 
the metal atoms and one of the carbon atoms in the $CP$-ring (top panel Fig.~\ref{2Dmag}).
This plane provides us a side view, with respect to the axis of the nanowire.
The lower panel of Fig. \ref{2Dmag} 
illustrates the ``second'' plane that has the edges parallel to the $x$ and $y$ axes and provides 
a top view of the spin density in the cyclopentane ring. For both spin orientations 
we use ten equidistant contours 
and an increment of $\pm$ 0.0025 $e/Bohr^3$ for the up / down spin densities.
Note that the spin density on carbon atom has a maximum value close to 0.0075 
$e/Bohr^3 $ (i.e. three contour lines). As seen on the contour plot, this 
maximum value is at least two orders of magnitude lower than the values obtained 
for the spin densities on the metal site. For Fe the maximum spin density is located 
around its atomic position and has a value of about 1.2 $e/Bohr^3$ (picture not shown here).
In addition for the Fe sites the difference of real-space spin densities 
has a positive sign (blue contours), while on the carbon sites the difference is negative 
signaling an anti-parallel orientation of a small magnetic moment with respect to Fe magnetic moment.  

For $CP-V$ a non-zero spin-density contribution is present on the $CP$-rings, which makes this 
material to be an almost half metallic compound. 
We also noticed that in the case of $CP-Co$, the majority 
spins accumulate on one side of the cyclopentane ring, 
therefore contributing to the anti-ferromagnetic properties obtained for $CP-Co$. 

\section{Multi-orbital Hubbard model for half-metallic {\it CP-Cr} chain}
\label{MUH}
\subsection{Model setup}
In order to discuss further the magnetic properties of the half-metallic $Cp-Cr$ metallocene 
nano-wire we supplement the DFT analysis with many-body calculations based on a
multi-orbital Hubbard model. Such calculations were performed recently for the
prototype half-metallic ferromagnet NiMnSb including local\cite{ch.ka.03} and
non-local correlation \cite{al.ch.08b,al.ch.10} effects, demonstrating the appearance
of significant many-body induced states within the half-metallic gap leading to
the disappearance of half-metalicity. Our primary goal is to investigate the existence
of similar effects in the half-metallic metalocene nano-wires, in particular the
stability of the half-metalicity in the presence of the Coulomb interaction.
In the present section we discuss the construction of the low energy Hamiltonian
in a similar way as has been recently performed for NiMnSb~\cite{al.ch.08b,al.ch.10}, 
CrO$_2$~\cite{ya.ch.06} or TiN~\cite{al.ch.09} in the framework of N-th 
order muffin-tin-orbital method. Alternative downfolding schemes, can be also 
formulated within tight-binding approaches \cite{ya.wa.09,ya.wa.11}.

\begin{figure}[h]
\begin{center}
\rotatebox{270}{\includegraphics[height=1.\columnwidth]{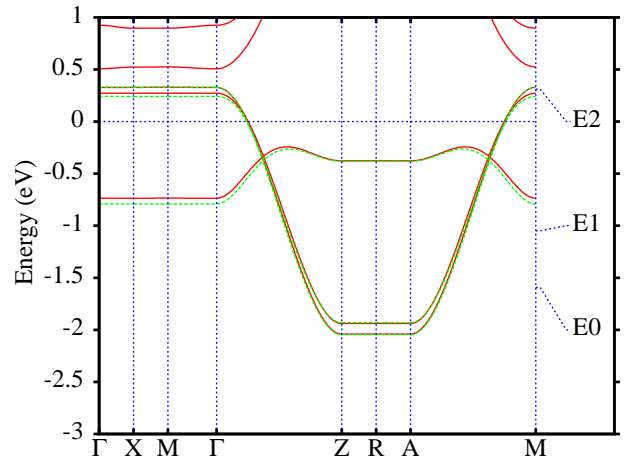}}
\end{center}
\caption{(color online) Band-structure of $CP-Cr$ around Fermi-energy obtained with the
LDA calculation (red line) for the complete orbital basis set and the NMTO bands
obtained after downfolding to the effective $xy$, $x^2-y^2$ and $z^2-1$ orbitals.}
\label{NMTO_bnds}
\end{figure}

The starting point in the combined electronic structure and many-body calculation
is the formulation of the low energy model Hamiltonian. Specifically, the uncorrelated
part of the Hamiltonian is obtained from the downfolding procedure \cite{an.sa.00,zu.je.05}
within the N-th order muffin-tin-orbital (NMTO) method.
The NMTO method~\cite{an.sa.00,zu.je.05} can be used to generate truly minimal basis sets
with a massive downfolding technique. Downfolding produces bands with a reduced basis which
follow exactly the bands obtained with the full basis set.
The truly minimal set of symmetrically orthonormalized NMTOs is a set of Wannier
functions. In the construction of the NMTO basis set the active channels are forced
to be localized onto the eigenchannel ${\bf R}lm$, therefore the NMTO basis set
is strongly localized.
For $Cp-Cr$ we choose to downfold (i.e. to 
integrate out) all orbitals except the Cr-3d ($xy, x^2-y^2$, ${z^2-1}$) manifold.
At this point is important to mention that the NMTO calculations has been performed
considering the structure shown in Fig. \ref{struc} with the parameters discussed
in Table \ref{tabel}. The calculations considered the tetragonal symmetry in
such a case the Cr-3d($xy, x^2-y^2$) orbitals are nearly degenerate, while in the 
computation using SIESTA a complete degeneracy is seen as the $D_{5h}$ structural symmetry
of the ligand field is considered. This difference turns out not to be essential
in considering the many-body effects presented here. 
Thus, the effective Hamiltonian is confined to a set of effective three 
orbitals of ($xy$, $x^2-y^2$, $z^2-1$)-type in a reduced window of energies. 
For optimizing the energy window with respect to the orbitals we chose the following 
expansion points with respect to the Fermi level: $E_\nu-E_F=0.31eV$, $-1.05eV$ and  
$-1.59eV$, where the Fermi energy has the absolute value $E_F=-3.44$eV. In 
Fig.~\ref{NMTO_bnds} we show the eigenvalues of the effective Hamiltonian 
along some high-symmetry directions in comparison with the full orbital basis. 

Fourier-transformation of the orthonormalized NMTO Hamiltonian,
$H^{\rm LDA}({\bf k})$, yields on-site energies and hopping integrals,
\begin{equation}
H^{LDA}_{{\bf 0}m',{\bf R}m} \equiv \left\langle \chi _{%
{\bf 0}\,m^{\prime }}^{\perp }\left\vert \mathcal{H}%
^{LDA}-\varepsilon _{F}\right\vert \chi _{{\bf R} \,m}^{\perp}\right\rangle
\equiv t_{m^{\prime },m}^{xyz},
\end{equation}
in a Wannier representation, with the orthonormal $ \vert \chi_{{\bf R} \,m}^{\perp}\rangle$-NMTO Wannier functions. 
$m$ is labeled by the three $xy$, $x^2-y^2$, $z^2-1$ orbitals. In this representation
the on-site matrix elements $t_{m^{\prime },m}^{000}$ are nearly diagonal, a very small 
non-zero coupling between the $x^2-y^2$ and  ${z^2-1}$ orbitals is obtained, which is 
a consequence of the considered tetragonal symmetry. A set of rotated orbitals has
to be introduced such that the local single-particle density matrix is diagonalized. Such
a procedure is frequently used~\cite{ya.wa.11} and in the present case this leads
to a small change in the effective  hopping parameters of $Cp-Cr$.
The on-site and the directional hopping matrix elements within the rotated basis-set 
are given (in units of eV) by:
\begin{eqnarray}
\label{mat_ham}
t_{m',m}^{000} &= \left(
\begin{array}{rrr}
-4.245 &    0  &    0   \\
   0  & -4.287 &    0   \\
   0  &    0  & -3.869
\end{array}
\right) \nonumber , \\
t_{m',m}^{00\pm1} &= \left(
\begin{array}{rrr}
 -0.545 &    0 &   0    \\
  0    & -0.589 & -0.02  \\
  0    & -0.02 & 0.091
\end{array}
\right) 
\end{eqnarray}
Thus, the non-interacting part of the effective Hamiltonian for
$CP-Cr$ has the form:
\begin{equation}\label{ho}
H_0 = \sum_{{\bf R'},{\bf R},\{m', m \}, \sigma} t_{m',m}^{{\bf R'}-{\bf R}}  c_{{\bf R'}m' \sigma}^\dag
c_{{\bf R}m \sigma} + h.c.  
\end{equation}
To take into account correlation effects to the non-interacting Hamiltonian in eq. \ref{ho} we add 
the Hubbard part such that  a 3-band ``correlated'' Hamiltonian is cast in the form:
\begin{eqnarray}
H &=& H_0 + \sum_{{\bf R}m}Un_{{\bf R}m\uparrow}n_{{\bf R}m\downarrow} \nonumber \\
&+&\sum_{{\bf R},m<m',\sigma,\sigma'}(U'-J\delta_{\sigma,\sigma'})n_{{\bf R}m\sigma}n_{{\bf R}m'\sigma'}
 \nonumber \\
&+&\sum_{{\bf R},m<m'} J_{mm'} c^\dagger_{{\bf R}m'\uparrow} c^\dagger_{{\bf R}m\downarrow} c_{{\bf R}m'
\downarrow}c^\dagger_{{\bf R}m\uparrow} + h.c.\nonumber \\
&+&\sum_{{\bf R},m<m'} J_{mm'} c^\dagger_{{\bf R}m'\uparrow} c^\dagger_{{\bf R}m'\downarrow} c_{{\bf R}m
\downarrow}c^\dagger_{{\bf R}m\uparrow} + h.c. 
\label{Mu-Hubb1}
\end{eqnarray}
where  $H_0$ is given by eq. (\ref{ho}), 
$n_{{\bf R}m\sigma}=c^\dagger_{{\bf R}m\sigma} c_{{\bf R}m\sigma}$
is the number of particle operator and $c^\dagger_{{\bf R}m\sigma}(c_{{\bf R}m\sigma})$ 
are the usual fermionic creation(annihilation) operators acting on an electron with spin 
$\sigma$ at a site {\bf R} in the orbital $m$.
The on-site Coulomb interactions are expressed in terms of two parameters $U$ and $J$
via: $U_{mm}=U$, $U_{mm'(\neq m)}=U'=U-2J$, $J_{mm'}=J$ \cite{li.ka.98}. In our calculations
we used values of $U$ in the range of $2-4eV$ and $J$ in the range of $0.6-1.2eV$ as they are
usual parameters for transition metal elements~\cite{mi.eb.05,br.mi.06,ch.mi.06}. 
Similar values were used to include correlation effects within mean-field GGA+U 
calculations \cite{we.oz.08,we.oz.09} in multiple-decker-type of compounds. Note 
that  corrections for a double-counting of the interaction within the manifold of 
orbitals considered, only produces an irrelevant constant shift of the chemical 
potential \cite{pa.bi.04,pe.ma.03} as we are considering a model Hamiltonian with 
fixed number of electrons.

\subsection{The correlated ferromagnetic state}
The total Hamiltonian in
eq. (\ref{Mu-Hubb1}) is solved by the VCA method, that is an extension of Cluster Perturbation
Theory (CPT)~\cite{gr.va.93,se.pe.00,ov.sa.89}.
In this approach, the original lattice is divided into a set of disconnected clusters and the
inter-cluster hopping terms are treated perturbatively. VCA additionally includes ``virtual''
single-particle terms to the cluster Hamiltonian, yielding a so-called reference system, and
then subtracts these terms perturbatively. The ``optimal'' value for these variational
parameters is determined in the framework of the Self-energy Functional Approach
(SFA)~\cite{pott.03,pott.03.se}, by requiring that the SFA grand-canonical potential $\Omega$
is stationary within this set of variational parameters.
\begin{figure}[h]
\includegraphics[width=0.99\columnwidth]{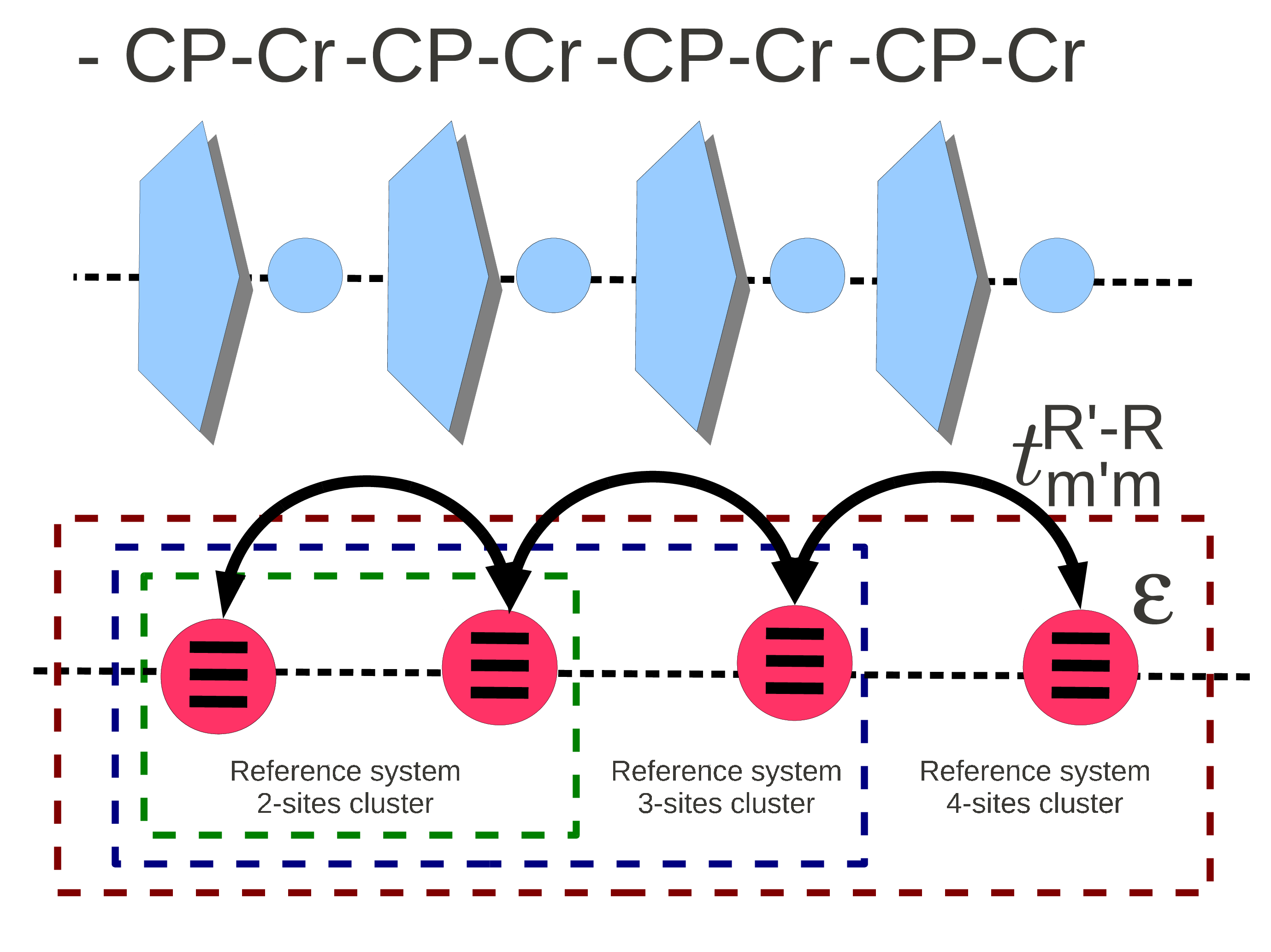}
\caption{(color online) Schematic representation for the model $CP-M$ chain.
Three different choices for the reference systems are shown: the 2,3 and the 4 sites clusters.
Intra- and inter-site couplings are described by the $t_{m',m}^{{\bf R'}-{\bf R}}$ matrix, with
$\epsilon$ the on-site energies for the effective d-orbitals rotated in the local basis.}
\label{chain_homo}
\end{figure}

In fig. \ref{chain_homo} we show the choice of the reference system, obtained by disconnecting 
the multiple decker nanowire (i.e. the chain of $CP-M$ units) into a set of sites  
connected by the same intra- and inter-cluster hopping $t_{m',m}^{{\bf R'}-{\bf R}}$. 
The electronic states connected to the $CP$-rings and some of the lower and higher 
$d-$ energies were integrated-out so that the electronic structure around the 
Fermi level is described by the low energy Hamiltonian Eq. \ref{Mu-Hubb1} formed 
with the three active effective orbitals. 

%
Our finite size scaling analysis has been performed for the fixed values of $U=2eV$ and 
$J=0.6eV$ parameters for the different clusters of two, three and four sites. The results
indicate that larger the cluster size convergence and optimization of the VCA grand potential
is still easily achieved. In particular, the self-consistent solutions for the grand potential, 
in the case of three and four sites clusters provides similar occupations to the non-interacting 
case.  The non-interacting results were obtained by self-consistent calculations of the grand 
potential in which the exchange splitting has been considered as variational
parameter parameter. For the four sites cluster, a total magnetic moment of $1.10 \mu_B$ is 
obtained within the non-interacting calculation while for the interacting case a value of 
about $0.95-0.97 \mu_B$ is obtained depending on the strength of average Coulomb parameters ($U$ and $J$).

\begin{figure}[htbp]
\includegraphics[width=0.99\columnwidth]{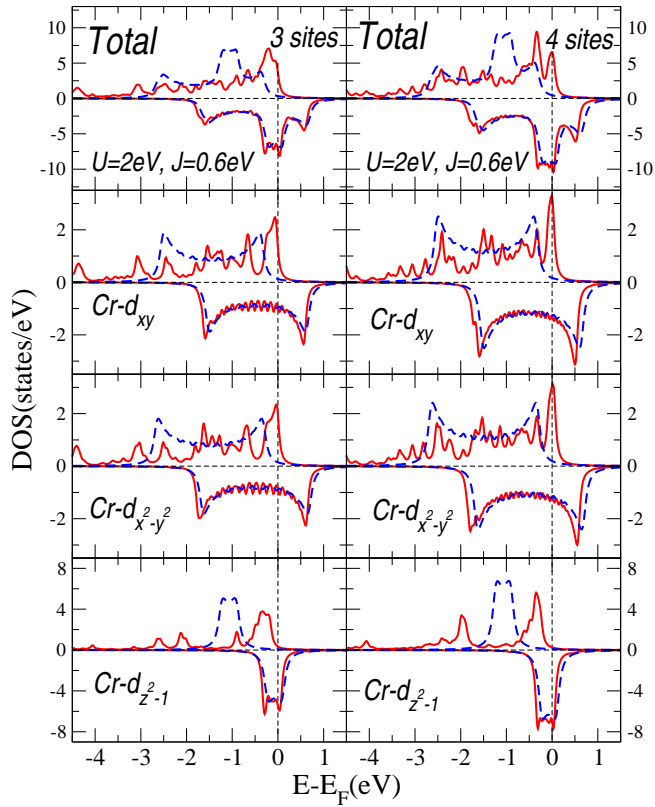}
\caption{(color online) Orbital resolved densities of states of $CP-Cr$ chain for 
the clusters containing three (left column) and four (right column) sites, and the 
average Coulomb and exchange interactions are $U=2eV$ and $J=0.6eV$. At the non-interacting
level (dashed-blue line) a half-metallic solution is obtained, while in the presence of 
interaction (solid-red line) majority spin gap is closed and a metallic solution is obtained.}
\label{vca_s_cr}
\end{figure}
In fig. \ref{vca_s_cr} we show the total and orbital resolved density of states,
for the clusters of three and four sites and fixed values $U=2eV$, $J=0.6eV$. 
As one can see, the majority spin gap is filled and normal ferromagnetic ground state is obtained
with a considerable spectral weight just below the Fermi level. In comparing to the non-correlated 
calculations (dashed-blue line), most significant changes are seen in the majority spin channel for 
all three orbitals. The majority $z^2-1$ orbital is pushed towards the Fermi level and splitted by 
electronic correlations, nevertheless it remains completely occupied. The $xy$ and $x^2-y^2$ orbitals 
remain almost degenerate and are shifted also towards the Fermi level. In addition they develop a 
double peak structure just below  Fermi level.  
Increasing the size of the cluster (see left- and right-column of Fig.\ref{vca_s_cr}), there is no 
significant change within the minority spin channel. All essential features of density 
of states remain unchanged when comparing the results of calculations for three and four sites clusters.
A close look reveals only the scale difference connected of the number of sites considered within the cluster. 

\begin{figure}[htbp]
\includegraphics[width=0.99\columnwidth]{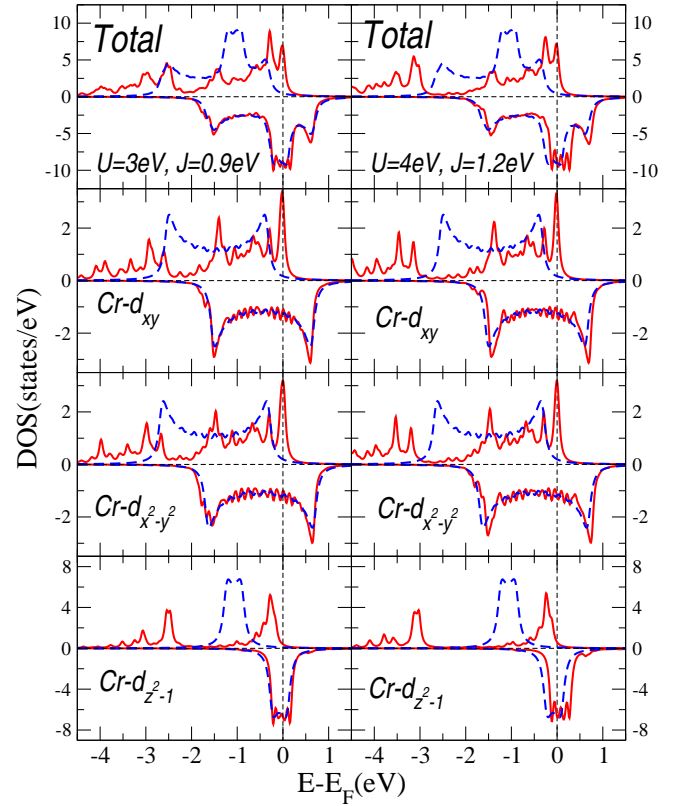}
\caption{(color online) Orbital resolved densities of states for the
four site cluster (interacting/non-interacting  - red solid/blue dashed line) 
with increasing values of $U$ and $J$ as indicated
in the figure.}
\label{vca_u_cr}
\end{figure}
\begin{figure}[htbp]
\includegraphics[width=0.99\columnwidth]{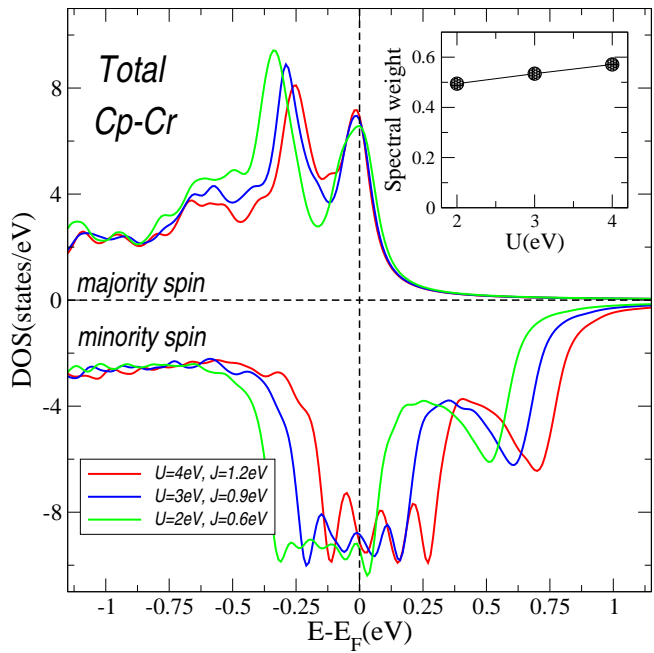}
\caption{(color online) Total density of states for the four site cluster
for different values of $U$ and $J$ parameters. Inset presents the integrated
spectral weight of the peak centered at $E_F$.}
\label{vca_ef_cr}
\end{figure}

In Fig. \ref{vca_u_cr} we present the results for the four site cluster, for two distinct
values of $U$ and $J$. 
Increasing the values of $U$ and $J$, the density of states features situated 
between $-4$ and $-3eV$ are further renormalized towards higher binding energies.
Within an energy window of $\pm 1eV$ around Fermi level (see Fig. \ref{vca_ef_cr})
there is a slight shift of minority spin states towards Fermi level, while for the 
majority spins, spectral weight transfer towards Fermi level  is slightly increased 
for larger U. For the majority electrons, DOS in the close vicinity of Fermi level 
obtained within the many body calculations shows the presence of some states
pinned almost at the Fermi level. The position of these pinned states remains practically unchanged, 
for the different $U$ and $J$ parameters, while their spectral weight is slightly 
increasing with $U$ and $J$. From the orbital projected densities of states seen in 
Fig. \ref{vca_s_cr} and Fig. \ref{vca_u_cr} we can identify  its composition 
as being mainly determined by the $xy$ and $x^2-y^2$ orbitals.
Around 0.25eV below Fermi level, the peak having mainly a $z^2-1$ character 
is seen to shift towards $E_F$ when $U$ and $J$ increase. In addition 
spectral weight is transferred towards the peak pinned at the Fermi level. 
Within the minority spin channel, increasing $U/J$ leads to a visible shift 
of the $z^2-1$ orbital towards and above the Fermi level. 
The inset 
of Fig.~\ref{vca_ef_cr} shows the integrated spectral weight of the states pinned
in the close vicinity of Fermi level. A similar effect has been recently discussed for the 
ferrimagnet Mn$_2$VAl in which the local but dynamic electronic correlations captured 
by DMFT~\cite{ch.ar.09} demonstrated the closure of the gap and a very strong
depolarization effect. For both these compounds LDA based DFT calculations predict
a half-metallic majority spin gap, but for both materials correlations effects are 
shown to destroy half-metalicity, by creating many-body induced states just below 
the Fermi level. Because of their position below Fermi level and their considerable 
spectral weight, one can expect that these states might be detectable by spin-polarized 
photoemission, which constitutes an interesting possibility to investigate many-body
effects in such organo-metallic spintronic materials.

\section{Conclusion}
In this work we studied the electronic structure and magnetic properties of 
a series of metallocene multiple-decker sandwich nanowires with the formula 
$(C_5H_5)_2M_2$, where M=Ti to Ni. Our results show that a broad variety of 
electronic and magnetic properties are predicted in these wires (i.e. insulator, 
metallic, half-metallic). Based on the results of the binding energies, we 
show that all the structures are energetically stable. 
We demonstrate that physical properties of the multiple-decker nanowires 
are determined by  structural relaxation. Depending on the transition-metal
element we found regular alternations of the geometric parameters such as 
bondlengths and cell length, that determines the magnetic properties of $(C_5H_5)_2M_2$ 
systems. 
Accordingly we found that $CP-Co$ is an anti-ferromagnetic insulator with a large gap 
of about $1.5eV$. The Mn moment's reduction to zero in $CP-Mn$ and the considerable 
turn-up of the V moment in $CP-V$ is obtained. For the $CP-V$ and $CP-Ti$ both ferro 
and antiferro solutions are possible which allows us to estimate the exchange couplings 
between the V-V and Ti-Ti atoms. These are of magnitude $J_{V}=0.34eV \approx 3945 K$ and 
$J_{Ti}=0.17eV \approx 1973K$ and the corresponding ground state in these cases is metallic 
ferro and anti-ferromagnetic respectively. In all calculations the $CP-$rings carry a
very small induced magnetic moment, that contributes into the stabilization of the 
structure and the transmission of the magnetic interactions through the wire. For the 
Fe-based and Cr-based sandwiches we found within the DFT - GGA calculations stable 
half-metallic solutions. 

For the $CP-Cr$ half-metallic wire we have constructed a low-energy Hamiltonian by
downfolding all orbitals except Cr-3d($xy$, $x^2-y^2$, $z^2-1$) orbitals which provides 
an effective low-energy model. This simplified model-Hamiltonian is used to investigate 
correlation effects in the half-metallic ground state. We have solved the three-band 
Hubbard Hamiltonian within the Variational Cluster Approach. The main effects determined
by electronic interactions are: (i) within the minority channel (spin down) the 
$z^2-1$  orbital is less occupied and more itinerant, while the other minority spin orbitals do not 
suffer significant change; (ii) for the majority electrons (spin up channel) many body state
of $xy$ and $x^2-y^2$ origin are pinned within the half-metallic gap just below 
Fermi level. Their position is practically unchanged for different values of $U$ and $J$, 
in contrast to the occupied $z^2-1$ orbitals that shifts towards the Fermi level, 
with increasing $U$. The overall conclusion is that the majority spin half-metallic ground 
state is not robust against the presence of local Coulomb correlations, and a correlated
ferromagnetic ground state is obtained in which the majority spin gap is closed
by many-body induced states. 

{\bf Acknowledgements}. We thank I. Leonov for useful discussions. 
The calculations were performed in the Datacenter of NIRDIMT. CM acknowledges the financial
support offered by the Augsburg Center for Innovative Technologies (ACIT), University of Augsburg,
Germany.

\bibliography{references_database}
\end{document}